\documentclass[12pt]{article}

\usepackage{mystyle} 

\usepackage{mathrsfs}  
\usepackage[T1]{fontenc} 
\usepackage{tikz,caption,subcaption} 
\usetikzlibrary{decorations.markings}

\definecolor{lightblue}{rgb}{.1,.4,.5}
\definecolor{brown1}{rgb}{.64,.43,.138}

\usepackage{epsfig,euscript,xspace, color}
\usepackage{mathtools}

\newcommand{\be}{\begin{equation}}
\newcommand{\ee}{\end{equation}}
\newcommand{\refb}[1]{(\ref{#1})}

\newcommand{\sgn}{\mathop{\mathrm{sgn}}}
\newcommand{\Pf}{\mathop{\mathrm{Pf}}}

\newcommand{\nn}{\ensuremath{\nonumber{}}}

\title{ Phase Thermalization: from Fermi Liquid to Incoherent Metal}

\abstract{When a system consists of a large subsystem (bath) and a small one (probe), thermalization implies induction of temperature of the bath onto the probe.
	If both the bath and the probe are described by same microscopic Hamiltonian, thermalisation further entails that the probe imbibes the phase of the bath. We refer to  this phenomenon as {\it phase thermalization}. 
	However, it is not clear whether this phenomenon is realizable when the probe and the bath are described by different microscopic Hamiltonians. We show {\it phase thermalization} is possible even when the microscopic Hamiltonians differ significantly. 
	We provide an explicit example, where the probe is a Fermi liquid realised by a Majorana chain with $n \gg 1$ fermions per site interacting through random hopping and the bath is an incoherent metal described by another Majorana chain with $N > n$ fermions per site interacting through arbitrarily long range random four-fermion interaction. In deep infrared (\emph{i.e.} at very low energies), the probe turns into an incoherent metal, with Lyapunov spectrum and diffusion coefficient identical to the bath. }

\author{Pinaki Banerjee$^{1,2,3}$, Bidisha Chakrabarty$^{4}$, Swapnamay Mondal$^{5,6}$
	\\~\\
	
		$^1$ {\it Institute of Physics, Sachivalaya Marg,
		Bhubaneswar, Odisha, 751005, India}  \\
		$^2${\it ~Homi Bhabha National Institute, Anushakti Nagar, Mumbai, 400085, India}
	\\
	$^3$ {\it School of Natural Sciences, Institute for Advanced Study, Princeton, New Jersey 08540, USA}
	\\
	$^4$ {\it University of Southampton,
		University Road,
		Southampton 
		SO17 1BJ,
		United Kingdom}
	\\
		$^5$ {\it Department of Physics, Institute of Science, Banaras Hindu University, Varanasi, 221005, India}\\
	
	$^6$ {\it School of Mathematics, Trinity College, Dublin 2, Ireland}
	\\~\\
		{\tt e-mails: 
		\href{mailto:pinaki@ias.edu}{\texttt{pinaki@ias.edu}}, 	\href{mailto:b.chakrabarty@soton.ac.uk}{\texttt{b.chakrabarty@soton.ac.uk}}, 	\href{mailto:swapno@maths.tcd.ie}{\texttt{swapno@maths.tcd.ie}}
	    }
}

\begin{document}

	\maketitle
	\flushbottom
	%
	
%%%%%%%%%%%%%%%%%%%%%%%%%%%%%%%%%%%%%%%%%%%%%%%%%%%%%%%%%%%%%%%%%%%%%%%%%%%%%%%%%%%%%%%	 	

\section{Introduction}  \label{s1}
One of the most ubiquitous natural phenomenon is thermalization \cite{Landau:1980mil, Srednicki_1994, Deutsch}. Classically it refers to phase space density attaining a universal form parameterised by an emergent quantity, namely temperature, which further takes the same value throughout the system considered \cite{Landau:1980mil}. 
Quantum mechanically, thermalization is understood in the framework of Eigenstate Thermalization Hypotheses \cite{Srednicki_1994, Deutsch}.
Although there exist counterexamples such as many body localization \cite{PhysRev.109.1492, PhysRevB.21.2366, PhysRevLett.78.2803, PhysRevLett.95.206603, BASKO20061126, Nandkishore:2014kca}, vast majority of physical systems exhibit thermalization.

When a large system (bath) is connected to a smaller system (probe), thermalization manifests itself through induction of temperature from the bath onto the probe. 
Recently some evidence has emerged that under less general circumstances, a bath can induce finer quantities, such as Lyapunov exponent onto a probe \cite{Mondal:2019gec}.
This triggers one's intrigue about more exotic possibilities. For example, can a bath induce macroscopic properties other than temperature, say some transport properties? Or at a finer level, can a bath change the effective degrees of freedom of a probe? 
Possibly most fascinating of all, {\it can a larger quantum many body system induce its phase onto a smaller one}? 

Occurrence like an ice cube melting in warm water, are {\it not} true examples of this phenomenon however. In such situations, the microscopic Hamiltonians describing the bath and the probe have the same form but due to difference in temperature, they undergo different  renormalization group (RG) flows leading to different macroscopic phases. 
The role of the bath is limited to inducing its temperature onto the probe and the rest is attributed to the dynamics of the probe itself. 
However, when the bath and the probe are described by microscopic Hamiltonians that are different in form, induction of phase can not be caused by induction of temperature. Hence phase induction in such circumstances presents a genuinely novel phenomenon, which we refer to as {\it phase thermalisation}.  

This is to be distinguished from proximity induced phase transitions \cite{PhysRevLett.100.096407,  Sitthison_2014,Tanaka_2009,Linder_2010,Black_Schaffer_2013,Black_Schaffer_2013_2,Tanaka_2012,Meng_2012,Lee_2014,Khanna_2014,
	PhysRevB.18.1076,RevModPhys.36.225,PhysRevB.65.014509,PhysRevB.77.184507,heersche2007bipolar,sacepe2011gate,zareapour2012proximity,Wang_2013}, since the relative size of the inducing system (e.g. a superconductor) and the induced system (e.g. a topological insulator) is not of key importance there. Also the phase induction happens only across the boundary, falling off exponentially in the bulk. 
On the contrary, in phase thermalization, relative size should be the determining factor and over time the entire probe is expected to develop the phase of the bath. Under which specific circumstances  phase thermalization can arise? Whereas we do not have a general recipe, one possibility might be to consider a bath-probe system, where the dynamics is largely dictated by symmetry.

When an infinite dimensional symmetry is broken to a finite dimensional one, there may be several exotic possibilities. One such system is SYK model \cite{Kitaev_talk,Maldacena:2016hyu, Polchinski:2016xgd}, which describes a quantum dot comprising large number of disordered Majorana fermions. This system develops an emergent time reparameterization symmetry in the very low energy (deep infrared) regime, which is both spontaneously and explicitly broken. This particular pattern of symmetry breaking suffices to determine the low energy effective action \cite{Maldacena:2016upp}.

The SYK system however lacks spatial extent. A more attractive possibility is presented by lattice of SYK dots \cite{Gu:2016oyy, Berkooz:2016cvq, Jian:2017unn, Song:2017pfw}. While retaining the essential physics of SYK model, these models make it possible to define transport properties.  Specifically, these models describes Incoherent Metal (IM) and saturate chaos bound. This motivates us to consider an IM bath. 

For the probe, we look for phases of matter where symmetry does not play a strong role. Simplest choice for such a phase is a Fermi Liquid (FL) \cite{Landau:1956zuh}. Our goal in this paper is to explore how the phase of the probe, which is a FL, is affected when we let it interact with an IM bath. We are able to solve this explicit example of an  interacting bath-probe system in a particular thermodynamic limit \emph{viz.} $N > n \gg 1$ and $N \to \infty$, where $N$ is the number of fermions per site of the bath and $n$ is the same for the probe. We find that the bath induces its phase on the probe and the probe is turned into an incoherent metal, when connected to the bath. Remarkably, as mentioned above, the interacting bath-probe system is solvable for any any size of the probe as long as $n< N$, \emph{i.e.} even when the probe is comparable to the bath in size. We also compute the  diffusion constant and Lyapunov spectrum of the system and find they are also induced onto the probe by the bath. Our conclusions remain unaltered under a large class of perturbations of the Hamiltonian over a fairly broad window of parameter space.

This paper is organized as follows. In Section \ref{sbath}, we introduce the IM bath. The bath is modelled by a one dimensional chain of SYK dots, with pairs of sites interacting through an arbitrarily long range four-Fermi interaction. The model is analytically solved in deep infrared, where the physics is dominated by Goldstones of spontaneously broken time reparameterization symmetry. Effective action for these Goldstones are determined. Diffusion coefficient, Lyapunov spectrum and butterfly velocity are computed from analysis of two points functions of these Goldstones. In Section \ref{sec3}, we discuss the probe and its interaction with the bath. Specifically, in subsection \ref{ssprobe} we introduce the FL probe, also modelled by a one dimensional chain of quantum dots, where each dot hosts $n<N$ Majorana fermions and different dots interact through random hopping. In subsection \ref{ssint}, the probe is connected with the bath through a random four-Fermi interaction, which is quadratic in both bath and probe fermions. Each probe site interacts with every other bath site, with interaction strength falling polynomially with distance. It is found that in the deep infrared, the symmetries and their spontaneous breaking thereof by the saddle point are upheld. We continue to integrate out the bath fluctuation around this saddle point to get an effective action for the probe exclusively. This action turns out to be dominated by same soft modes, consequently the probe imbibes the phase of the bath. In section \ref{sdisc}, we comment on some aspects of our main results and also mention some  experimental setups where phase thermalization can possibly be realized.  

%%%%%%%%%%%%%%%%%%%%%%%%%%%%%%%%%%%%%%%%%%%%%%%%%%%%%%%%%%%%%%%%%%%%%%%%%%%%%%%%%%%%%%

\section{Incoherent Metallic bath} \label{sbath}
IM-s \cite{Hartnoll:2014lpa} are particular kind of strange metals 
which characteristically  have high (linear with temperature) resistivity. Due to absence of quasiparticles, strange metals can not be described by otherwise extremely successful framework of Fermi-liquid (FL) theory. Moreover in IM even the momentum is not conserved (i.e. no overlap between momentum and conserved current) and is degraded very fast. This may be caused by strong disorder or large lattice scattering. Due to disorders/impurities there is no sound mode but only diffusion modes \cite{Hartnoll:2014lpa}. They can have very large resistivity (violate Mott-Ioffe-Regel (MIR) bound) \cite{Pakhira} and hence are examples of bad metals. Whereas systems with quasi particles (e.g. usual metals) in presence of disorder go to localized phase at some critical density, systems without quasi particles (e.g. holographic metals) at strong disorder (at fixed temperature) become IM-s. 

\begin{figure}
	\includegraphics[width=155 mm]{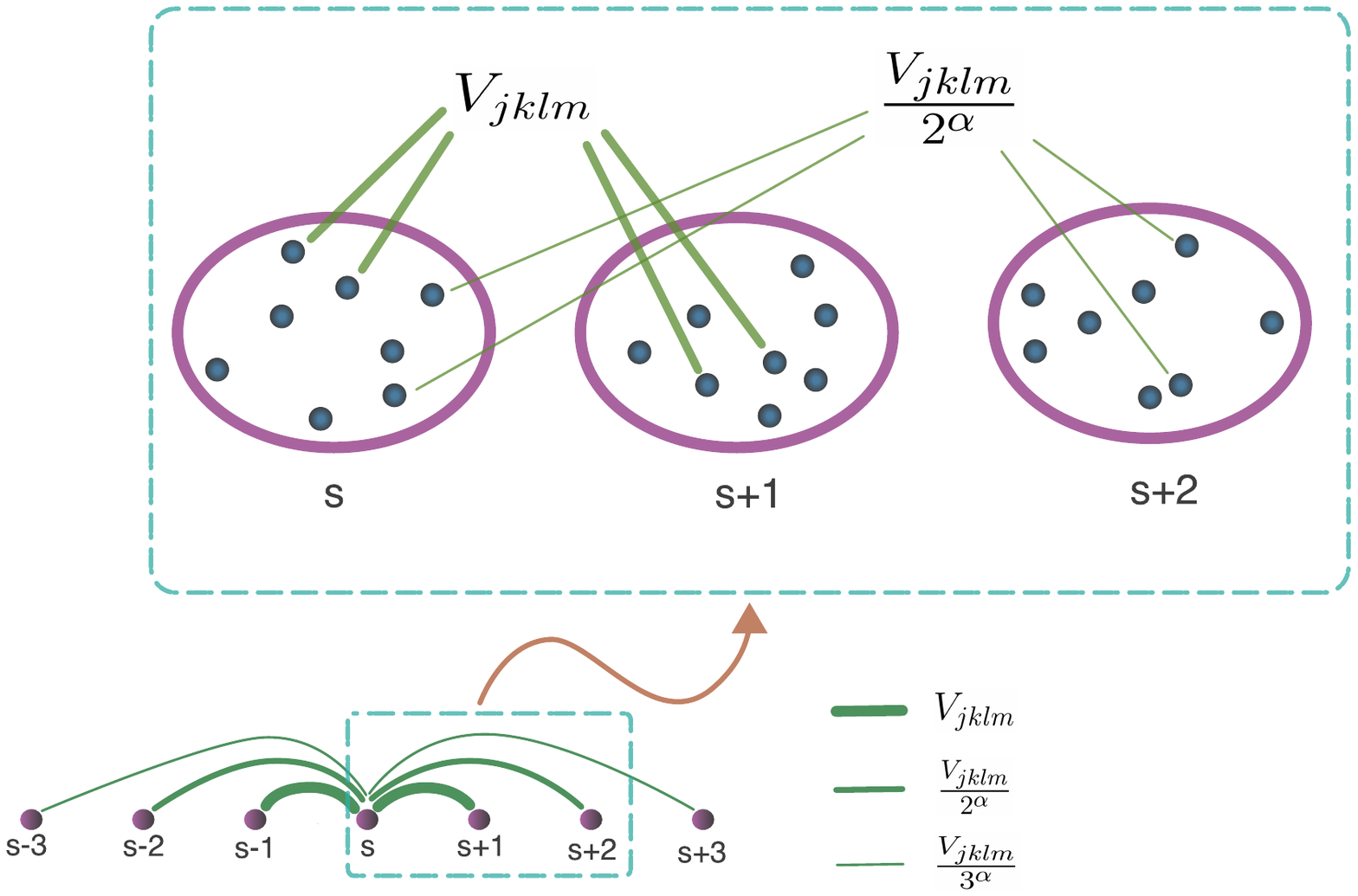}
	\caption{\textit{A pictorial representation of the lattice SYK system, described by the Hamiltonian \refb{hbath1d}. Each purple dot represents an SYK dot. Each such dot hosts $N \gg 1$ number of Majorana fermions, represented as blue dots in the inset. Green lines connecting SYK-dots represent pairwise interaction, and thinning of the green lines with distance of pairs represents decay of interaction strength. The strength of the interactions depend on the value of the parameter $\alpha$. The interaction becomes more and more short range for larger values of $\alpha$.}
	}\label{fig:syk-lattice} 
\end{figure}

Although finding explicit quantum matter realisation of IM-s has proven challenging, it is relatively easy to get IM-s in holographic models \cite{Vegh:2013sk, Kim_2014, Davison:2014lua, Davison:2015taa, Davison:2015bea, Lucas:2015lna}. E.g higher dimensional Sachdev-Ye-Kitaev (SYK) lattice models  \cite{Gu:2016oyy, Berkooz:2016cvq, Jian:2017unn, Song:2017pfw} are systems where one can realize them. They show diffusion of energy and number density even at zero temperature \cite{Davison:2016ngz, Blake:2016jnn}. 

We model the IM bath as a one dimensional chain of SYK-like dots. SYK model is a model of $N \gg 1$ number of disordered Majorana fermions with all to all interactions \cite{Kitaev_talk,Maldacena:2016hyu, Polchinski:2016xgd}. This model describes a non-Fermi liquid state that saturates the chaos bound \cite{Maldacena:2015waa} and exhibits emergent reparameterization symmetry, which is further broken spontaneously as well as explicitly. 
In recent years variants of SYK and SYK-lattice models \cite{Bi:2017yvx,Chen:2017dav,Jian:2017unn,Jian:2017jfl, banerjee2017solvable, Haldar:2019slc} have been studied extensively. This section will mostly be a review of the model of IM  bath introduced in \cite{Mondal:2018xwy}.

In the model considered, the dots interact in a pairwise manner through the Hamiltonian (see fig. \ref{fig:syk-lattice})
\begin{align}
	H_{bath} &= 
	\sum_{x, x' = - \infty , \atop{ x' >x}}^\infty \sum_{1 \leq j < k \leq N; \atop{1 \leq l < m \leq N}} \frac{V^{(xx')}_{jklm} }{(x'-x)^\alpha} \chi^x_j \chi^x_k \chi^{x'}_l \chi^{x'}_m \, ,
	\label{hbath1d} 
\end{align}
 where $V^{(xx')}_{jklm}$-s are random couplings obeying, $$\overline{ V^{(xx')}_{jklm} } = 0, \quad \overline{ V^{(x_1 x'_1)}_{jklm} V^{(x_2 x'_2)}_{j'k'l'm'} } = \frac{V_0^2}{N^3} \delta_{x_1 x_2} \delta_{x'_1 x'_2}  \delta_{jj'} \delta_{kk'} \delta_{ll'} \delta_{mm'},$$
with the overline denoting disorder average. The $\chi$-s are Majorana fermions and we are considering bi-local interactions that are located at lattice points $x$ and $x'$ and summing over all such interactions over the lattice. The lower indices of $\chi$ represent different flavours of the fermions. The parameter $\alpha$ dictates the strength of the interaction -- the interaction becomes more and more short range as $\alpha$ takes larger values. 
For SYK-dots it is natural to have an on site random quartic term : $\chi^x_j \chi^x_k \chi^{x}_l \chi^{x}_m$ (see e.g \cite{Gu:2016oyy,Song:2017pfw}). But this term has the effect of shifting $V_0^2$ by an additive constant. Hence we have dropped this term for the sake of simplicity. 

A distinguishing feature of the model \refb{hbath1d} is arbitrarily long range interaction. A model with nearest neighbor interaction is recovered in $\alpha \rightarrow \infty$ limit and describes an IM \cite{Gu:2016oyy}.
At finite $\alpha$, the system continues to be an IM until $\alpha =1$, where it undergoes a  change of phase \cite{Mondal:2018xwy}. We restrict ourselves to  $\alpha >1$, to stay away from any such pathology.

A random hopping being a relevant interaction, steers the system away to Heavy Fermi Liquid phase \cite{Song:2017pfw}. However this change occurs below the critical temperature  $T_c \sim \frac{s_0^2}{V_0 \sqrt{\zeta (2\alpha)}}$, with $s_0$ being the hopping strength. We choose to work above $T_c$, where hopping can be ignored without much consequence.
Interaction with phonons is expected not to affect the IM phase \cite{Guo:2019csw}.

Upon performing the disorder average, one finds that the effective action \refb{Sbath} depends on the fermions only through the bilocal bilinear combinations $\widetilde{G}^x(\tau, \tau') := \frac{1}{N} \sum_{i=1}^N \chi^x_i(\tau) \chi^x_i(\tau')$. It is useful to define $\widetilde{G}^x$ as new field of the theory along with the corresponding bilocal Lagrange multiplier field $\widetilde{\Sigma}^x$ which imposes the constraint $\widetilde{G}^x(\tau, \tau') = \frac{1}{N} \sum_{i=1}^N \chi^x_i(\tau) \chi^x_i(\tau')$ (see e.g. \cite{Gu:2016oyy}).

Solving for two point functions of $\chi$ fields to leading order in $1/N$, amounts to computing the saddle point values of $\widetilde{G}$ fields. 
In the deep infrared region, the saddle point equations of $\widetilde{G}$ and $\widetilde{\Sigma}$ together imply 
\begin{align}
	\frac{V_0^2}{2} \, \widetilde{G}^x \circ \left[ 
	\sum_{x' \neq x } \frac{1}{|x'-x|^{2\alpha}} \widetilde{G}^x \left( \widetilde{G}^{x'} \right)^2
	\right] &= -1 \, . \label{SDbath}
\end{align}
Here $\circ$ denotes convolution product, e.g. $( \widetilde{G}^x \circ \widetilde{G}^{x'}) (\tau_1, \tau_2) := \int d\tau'  \widetilde{G}^x (\tau_1, \tau') \widetilde{G}^{x'} (\tau', \tau_2) $. 

Remarkably,  \refb{SDbath} is invariant under an emergent time reparameterization symmetry
\begin{align}
	\tau &\rightarrow f(\tau) \, , \, 
	\widetilde{G}^x(\tau, \tau') \rightarrow |f'(\tau) f'(\tau')|^{-1/4} \widetilde{G}^x(\tau, \tau') \, \label{rep} .
\end{align}
We work in Euclidean time.
Assuming spatial translational invariance, the solution to  \refb{SDbath} is given by 
\begin{align}
	\nn
\widetilde{G}^x = \,	G_s(\tau_1, \tau_2) &:= \left( \frac{1}{4 \pi V_0^2 \zeta(2\alpha) } \right)^{1/4} \left[ \frac{\pi}{\beta \sin \frac{\pi \tau_{12} }{\beta}} \right]^{1/2} \sgn{} \tau_{12} \, ,\\
\widetilde{\Sigma}^x = \, \Sigma_s (\tau_1, \tau_2) &:=  V_0^2 \zeta(2\alpha) G_s(\tau_1, \tau_2)^3 \, , \label{Gsaddle1}
\end{align}
where $\tau_{12} = \tau_1 - \tau_2$, $\beta$ is the inverse temperature and  $\zeta(2\alpha) = \sum_{k \in \mathbb{N}} \frac{1}{k^{2\alpha}}$ is the Riemann zeta function. Since $G_s(\tau_1, \tau_2)$ depends only on $\tau_{12}$, in the following we often abbreviate $G_s(\tau_1, \tau_2)$ as $G_s(\tau_{12})$ and likewise for $\Sigma_s$.
The appearance of the coupling $V_0$ in the denominator and the scaling dimension of fermions being 1/4, both carry testimony to a strongly interacting phase.
The solution \refb{Gsaddle1} spontaneously breaks the reparameterization symmetry \refb{rep} down to a $SL(2, \mathbb{R})$ subgroup consisting of the global reparameterizations. 

Coming to transport, since the fermions are uncharged, only transport is of interest is that of energy.
The $\chi$ fermions themselves are localized at lattice sites and the bilocal fields are responsible for transport. Hence we look at the connected part of the ``gauge invariant'' (i.e. internal indices are summed over) four point functions of $\chi$ \cite{Gu:2016oyy}
\begin{align}
	\nn
	\frac{1}{N} \mathcal{F}_{xx'} (\tau_1, \tau_2; \tau_3, \tau_4) := &
	|G_s(\tau_1, \tau_2)|^{-1} |G_s(\tau_3, \tau_4)|^{-1}\\
	& \langle g^x(\tau_1, \tau_2) \, g^{x'} (\tau_3, \tau_4) \rangle \, , \label{Fcon}
\end{align}
where $|G_s|^{-1} g^x$ is the fluctuation of $\widetilde{G}^x$ around the saddle  \refb{Gsaddle1}. 
%}

Two point function of $g$-fields is computed from the momentum space action \refb{eqnA4}
\begin{align}
\nn
&S_{eff} = \frac{N}{8\pi} \int_0^{2\pi} dp \, g (p) \circ R(p) \circ g(-p) \, , \\
\nn
\text{with} \, \, \, &R(p) = 3 V_0^2 \zeta(2\alpha) \left( K_c^{-1} -1 \right) \\
&+ V_0^2 \left[ 2 \zeta (2\alpha) - Li_{2\alpha} (e^{ip})  - Li_{2\alpha} (e^{-ip}) \right] \, , \label{seffg}
\end{align} 
where $Li_n (z) := \sum_{k = 1}^\infty \frac{z^k}{k^n}$ is the polylogarithm function.  
The symmetric kernel 
\begin{align}
\nn
K_c (\tau_1, \tau_2; \tau_3, \tau_4) = &-3 V_0^2 \zeta(2\alpha) |G_s(\tau_1, \tau_2)|  G_s(\tau_1, \tau_3) \\
& G_s(\tau_2, \tau_4) |G_s(\tau_3, \tau_4)| \,  \label{kcdefn}
\end{align}
commutes with $SL(2, \mathbb{R})$ generators, making $S_{eff}$ in \refb{seffg} $SL(2, \mathbb{R})$ invariant.

Since \refb{seffg} is the action for fluctuations around a vacuum with spontaneously broken symmetry, it is expected to vanish when $g(p)$ stands for the Goldstones. This is made possible by the fact that $K_c$ has a unit eigenvalue. Goldstones are therefore identified with $p=0$ and $K_c=1$ modes. These modes however lead to a divergent $ \mathcal{F}$.
The cure to this problem is to note that the reparameterization symmetry itself was a consequence of restricting to deep infrared, or equivalently strong coupling regime $\beta \mathcal{V} \gg 1$, with $\mathcal{V}=V_0 \sqrt{2\zeta(2\alpha)}$. 
Taking $\mathcal{O}(1/\beta \mathcal{V})$ effects into account, which amounts to breaking the reparameterization symmetry explicitly, results in shift of the eigenspectrum of $K_c$, especially the unit eigenvalue is corrected as
$k(2,m) = 1 - \frac{\alpha_K |m|}{\beta \mathcal{V}} + \dots \, , $
where $\alpha_K\sim 2.85$ is a numerical constant and $m$ indexes Matsubara modes \cite{Maldacena:2016hyu}. Corresponding expansion of a soft mode is 

\begin{align}
	g(p; \tau_1, \tau_2) &= \sum_{|m| \geq 2} \varepsilon_m(p) g_m (\tau_1, \tau_2) \, , \label{gpfourier}
\end{align}
where
\begin{align}
	g_m (\tau_1, \tau_2) &= \frac{i \sqrt{\pi V_0^2 \zeta(2\alpha)}}{2}  \left( \frac{2\pi}{\beta} \right)^2 \frac{e^{-2\pi i m y_{12} /\beta}}{\sin \frac{\pi \tau_{12}}{\beta}} f_m(\tau_{12}) \, , \label{eqn8}
\end{align}
with $f_n(\tau_{12}) = -n \cos \frac{\pi n \tau_{12} }{\beta} + \frac{\sin \frac{\pi n \tau_{12}}{\beta}}{\tan \frac{\pi \tau_{12}}{\beta}}$ , $y_{12} = \frac{\tau_1 + \tau_2}{2}$.
This results in the following $\mathcal{O}(1/\beta \mathcal{V})$ action
(see appendix \ref{appA})

\begin{align}
	S_{eff}^{soft} 
	&=
	\frac{N \alpha_K}{ 256 \pi \mathcal{V}} \left( \frac{2\pi}{\beta}\right)^2 \int_0^{2\pi} dp \, \sum_{|m| \geq 2} \varepsilon_m(p) \varepsilon_{-m}(-p) |m| (m^2-1) \nonumber \\
	& \qquad  \qquad \qquad \qquad  \qquad \qquad \qquad \bigg[ 
	\frac{ 2\pi |m|}{\beta} + \frac{2\pi \mathcal{V}}{\alpha_K} \times \frac{2 \zeta(2\alpha) - Li_{2\alpha}(e^{ip}) - Li_{2\alpha}(e^{-ip}) }{3 \zeta(2\alpha)}
	\bigg] \, .
	\label{ssoftb}
\end{align}

In long wavelength limit, \refb{ssoftb} reduces to 

	\begin{align}
		S_{soft}^{bath} &\sim \frac{N \alpha_K}{256 \pi \mathcal{V}} \left( \frac{2\pi}{\beta}\right)^2 \int_0^{2\pi} dp \, \sum_{|m| \geq 2} \varepsilon_m(p) \varepsilon_{-m}(p) |m| (m^2 -1) \left[ \frac{2\pi |m|}{\beta} + \frac{2\pi \mathcal{V} \zeta(2\alpha-1)}{3 \alpha_K \zeta(2\alpha)} p^2 \right] \, .
		\label{ssoftbpsmall}
	\end{align}

This small, yet non-zero action\footnote{$S_{soft}^{bath}$ is really a misnomer since strictly speaking only $p=0$ modes are the soft modes. They appear as part of a continuum in momentum space, hence we extend our action to include momentum modes in a window around $p=0$. This allows us to capture the long wavelength physics.}
is a hallmark of this particular pattern of symmetry breaking \cite{Maldacena:2016upp}, and implies that the seeming Goldstones are merely  ``pseudo-Goldstones''\footnote{Pseudo-Goldstone is a standard terminology used for Goldstone modes corresponding to a broken symmetry that was not an exact (but only an approximate) symmetry to start with.}.
This action in turn leads to a $\mathcal{O}(\beta \mathcal{V})$ contribution to $\mathcal{F}$, which we shall refer to as $\mathcal{F}^{big}$, given by \refb{eqnA11}

	\begin{align}
		\frac{\mathcal{F}^{big}(p;\tau_1, \tau_2;\tau_3,\tau_4)}{G_s(\tau_{12}) G_s(\tau_{34})} &=\frac{32 \pi V_0\sqrt{2 \zeta(2 \alpha)}}{\alpha_K} 
		\sum_{|m| \geq 2} \frac{1}{|m| (m^2-1)} \frac{ e^{-2\pi i (y_{12} - y_{34})/\beta}}{|\omega_m| + D_{\alpha} p^2} f_m(\tau_{12})  f_m(\tau_{34}) \, , 
		\label{Fbigbath}
	\end{align}

where 
and $\omega_m= \frac{2\pi m}{\beta}$ are the Matsubara frequencies. $\mathcal{F}^{big}$ is parametrically larger than contributions coming from non-Goldstones.
%We emphasise that this form of $\mathcal{F}^{big}$ is entirely dictated by the pattern of symmetry breaking.
$\mathcal{F}^{big}$ is derived from $S^{bath}_{soft}$, which in turn is entirely dictated by the pattern of symmetry breaking, as shown in appendix \ref{appA}. Thus $\mathcal{F}^{big}$ too is entirely dictated by the pattern of symmetry breaking. E.g. each $m$ in the sum \eqref{Fbigbath} corresponds to a reparametrization which alters the mean field solution but not the  Dyson-Schwinger  equation \eqref{SDbath}. The modes $m=0, \pm 1$ are excluded because these corresponds to the reparametrizations preserved by the mean field solution, hence they do not represent (pseudo)-Goldstone modes.

The diffusion coefficient $D_\alpha$ is given by
\begin{align}
	D_\alpha &= 
	\frac{2 \sqrt{2} \pi V_0}{3 \alpha_K \sqrt{\zeta(2\alpha)}} \zeta (2\alpha -1) \, . \label{diffbath}
\end{align}
This saturates the bound conjectured in \cite{hartnoll2015theory}. Similar behaviour was found in certain holographic theories \cite{Blake:2016wvh,Blake:2016sud}. 

In the limit $\alpha \rightarrow \infty$, diffusion coefficient approaches the limit value $D_\infty := \frac{2 \sqrt{2} \pi V_0}{3 \alpha_K}$.  As one lowers $\alpha$, the value of $D_\alpha$ increases and finally diverges as $\alpha \rightarrow 1_+$. In fact around $\alpha=1$, $\zeta(2\alpha) - \frac{Li_{2\alpha} (e^{ip}) + Li_{2\alpha} (e^{-ip}) }{2}$ no more admits a Taylor expansion around $p=0$. More specifically one encounters a $p^2 \ln{} |p|$ term. The apparent divergence of $D_\alpha$ at $\alpha=1$ is same as the divergence of $\ln{} p^2$ as $p \rightarrow 0$. Note that $p^2 \ln{} |p|$ however is well defined as $p \rightarrow 0$. To summarize, in this regime diffusion coefficient is no more a meaningful concept. For $\alpha <1$, the $p^2$ term is replaced by $|p|^{2\alpha -1}$.The system  undergoes a change of phase at $\alpha=1$, which might be a crossover since no obvious symmetry is broken.
Finally at $\alpha=1/2$, the theory becomes ill defined due to divergence of $\zeta(2\alpha)$. 
We restrict to large $\alpha$ regime.

\begin{figure}%
	\centering
	\subfloat[\centering For  $p = 0.01$ ]{{\includegraphics[width=7.8cm]{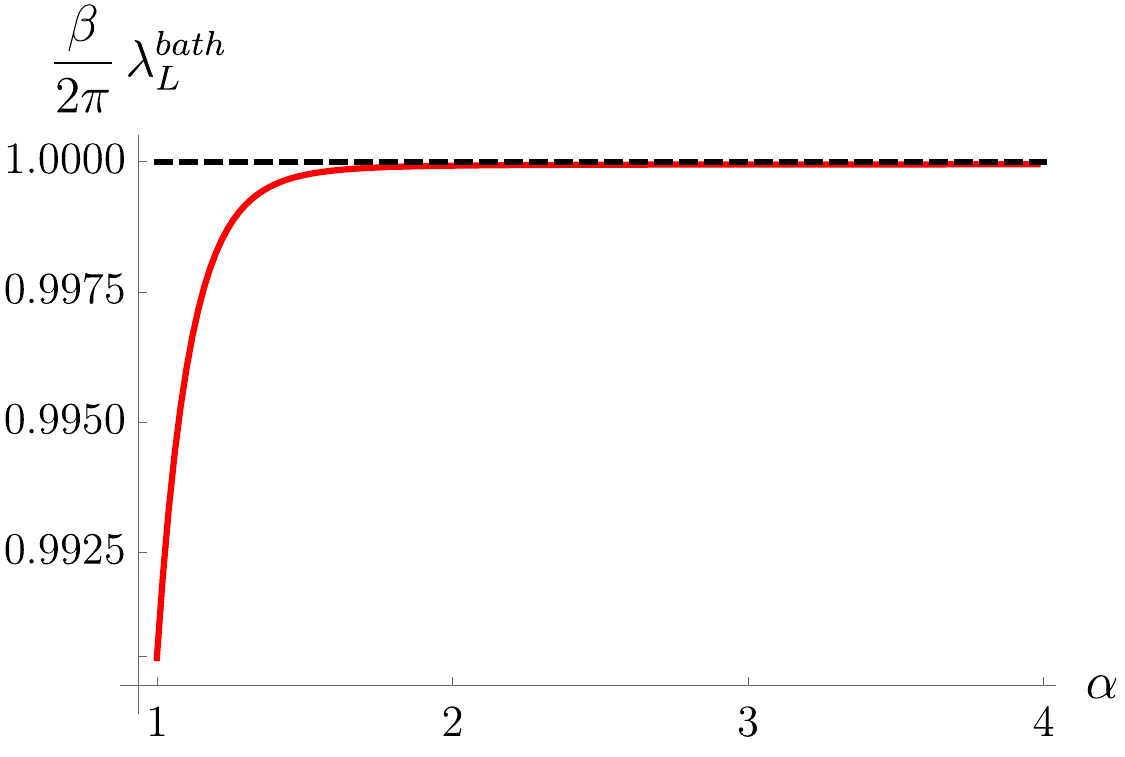} }}%
	\qquad
	\subfloat[\centering For $\alpha = 1.47$]{{\includegraphics[width=7.8cm]{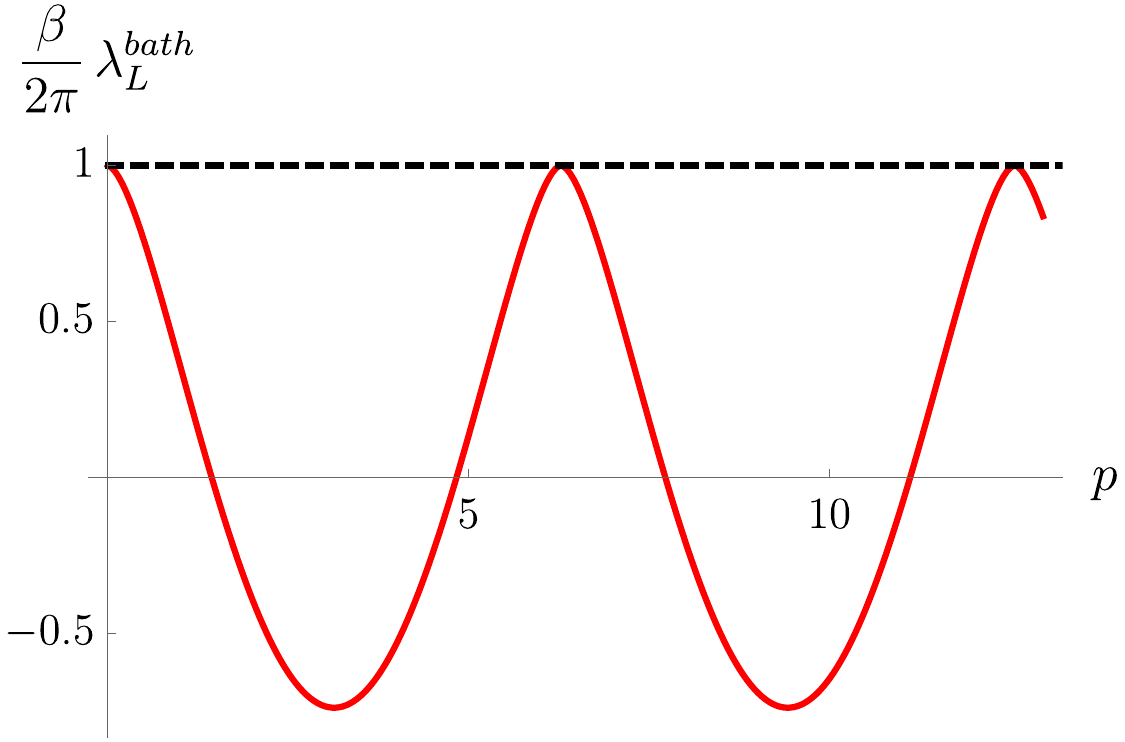} }}%
	\caption{\textit{The Lyapunov exponent \eqref{LyapunovExp} is plotted numerically using the analytically continued expression for $\text{Li}_s(z)$ as given by \eqref{analytic-cont-Li}. }}%
	\label{fig:Lyapunov}%
\end{figure}

Being bereft of quasi-particles, IM-s lack important scales such as quasiparticle mean free time and Fermi velocity. 
Lyapunov time $\tau_L=1/\lambda_L$ and butterfly velocity $v_B = \sqrt{2\pi D_\alpha/\beta}$ \cite{blake2016universal}, the velocity with which a chaos spreads spatially, have been suggested to play these roles respectively \cite{Blake:2016wvh, Blake:2016sud,Blake:2017qgd}.
In ``semi-classically'' early times,  chaos is captured in exponential early time growth of out-of-time-order correlation functions, or OTOC-s \cite{Larkin-Ovchinnikov,Maldacena:2015waa} 
\begin{align}
	%F(x,t) = 
	\frac{1}{N} \sum_{j,k=1}^N \langle \chi_{j,x}(t+ 3i \beta/4) \chi_{k,0}(i \beta/2) \chi_{j,x} (t + i \beta/4) \chi_{k,0} (0)  \rangle \, ,
\end{align}
which behave as $G(\beta/2)^2 \left( 1 - \mathcal{N} e^{\lambda_L t} \right)$ in early times, 
with $\mathcal{N} \sim 1/N$, and $\lambda_L$ being the Lyapunov exponent. 
Explicit computations give
\begin{align} \label{LyapunovExp}
	\lambda_L^{bath}(p) 
	&=
	\frac{2\pi}{\beta}  \frac{Li_{2\alpha} (e^{ip}) + Li_{2\alpha} (e^{-ip}) }{2 \zeta(2\alpha)}  \, .
\end{align}
Lyapunov spectrum depends on $\alpha$. Notice that  for $p \to 0$  limit $\lambda_L^{bath}(p)$ takes the maximal value $\frac{2\pi}{\beta}$.  Thus the spatially constant mode ($p \to 0$) always saturates the chaos bound \cite{Maldacena:2015waa}.  As explained above,  our system remains in an Incoherent metallic phase for  $\alpha >1$ and  strictly speaking, our results should be trusted  for $\alpha > 1$.  Assuming \eqref{LyapunovExp} has a valid analytic continuation over the complex $\alpha$ plane\footnote{The expression \eqref{analytic-cont-Li} is valid for all complex values of $s$. Here is how it's defined using Lerch transcendent (see \href{https://dlmf.nist.gov/25.14}{25.14} of \cite{NIST:DLMF})\begin{align}
	\Phi(z, s, a) = \sum_{k=0}^{\infty} \frac{z^k}{(a + k)^s} = \frac{1}{2 a^s} + \int_{0}^{\infty} \frac{z^t}{(a+t)^s}  - 2 \int_{0}^{\infty} \frac{\sin(t \log z - s \tan^{-1}(t/a))}{(a^2 + t^2)^{s/2}(e^{2 \pi t } -1)} \, d t
	\end{align}
\begin{align}
\text{Li}_s(z) \equiv z \, \Phi(z, s, 1) = \frac{z}{2} + \int_0^{\infty } \frac{z \sin \left(s \tan ^{-1}(t)-t \log (z)\right)}{\left(t^2+1\right)^{s/2} \sinh (\pi  t)} \, d t \, .
\end{align}} in terms of the integral representation of the $Li_{2 \alpha} (e^{\pm i p})$,
	\begin{align} \label{analytic-cont-Li}
\text{Li}_s(z)= \frac{z}{2} + \int_0^{\infty } \frac{z \sin \left(s \tan ^{-1}(t)-t \log (z)\right)}{\left(t^2+1\right)^{s/2} \sinh (\pi  t)} \, d t,  
	\end{align}
we can explore different regime in $\alpha$ and $p$. In $\alpha \rightarrow \infty$ limit, $\lambda_L^{bath}(p)$ reduces to $\frac{2\pi}{\beta} \cos p$ (see Figure \ref{fig:Lyapunov}). \\

\begin{figure}
		\centering
	\includegraphics[width=80 mm]{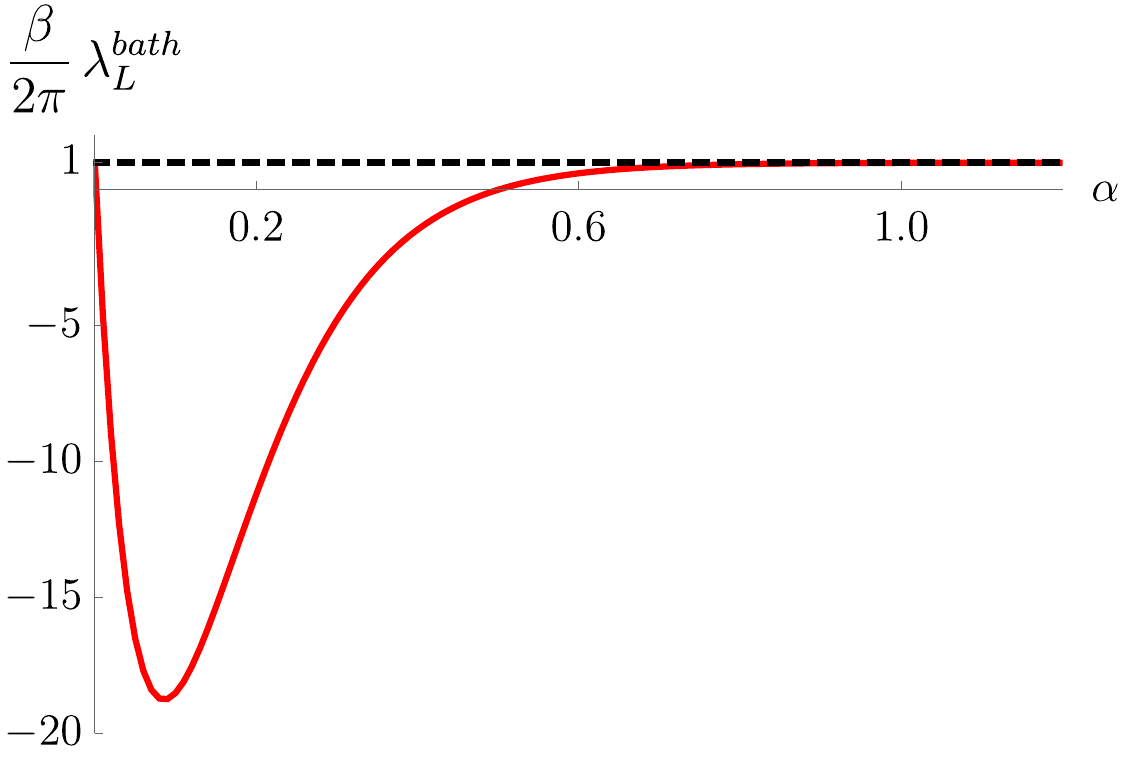}
	\caption{\textit{Strictly speaking we should restrict ourselves to $\alpha > 1$ but if we plot the Lyapunov exponent for all positive $\alpha$, the behaviour looks smooth even for $\alpha < 1$.  The Lyapunov exponent is negative for smaller values of $\alpha$. Potential hints of some interesting interplay between conservative ($\lambda_L^{bath} = 0$),  dissipative ($\lambda_L^{bath} < 0$) and chaotic ($\lambda_L^{bath} >0$) dynamics}}
	\label{fig:lambda-for-alpha-less-than-one} 
\end{figure}

Here we make a speculative but intriguing observation. We have emphasized before that we restrict ourselves to $\alpha > 1$ to avoid a potential singular behaviour at  $\alpha = 1$. But with the help of \eqref{analytic-cont-Li}, we can take formal limit $\alpha \to 0$ which gives $ \lambda _L^{bath} \simeq \frac{\beta}{2 \pi } \big(1-\frac{2 \pi  \alpha }{p} \big)$. We can also numerically plot the Lyapunov exponent for $\alpha < 1$ and obtain Figure \ref{fig:lambda-for-alpha-less-than-one}.  Note that if  the  expression for Lyapunov exponent remains valid for any positive values of $\alpha$, there can be an intriguing interplay between conservative/steady state ($\lambda_L^{bath} = 0$),  dissipative/stable ($\lambda_L^{bath} < 0$) and chaotic ($\lambda_L^{bath} >0$) dynamics in the considered system. Understanding this interesting phenomenon is out of the scope of the present paper and asks for further future explorations.

%%%%%%%%%%%%%%%%%%%% %%%%%%%%%%%%%%%%%%%% %%%%%%%%%%%%%%%%%%%% 
\section{Probe and bath-probe interaction}\label{sec3}

%%%%%%%%%%%%%%%%%%%% 
\subsection{Free Fermi liquid probe}\label{ssprobe}
\begin{figure}
	\includegraphics[width=170mm]{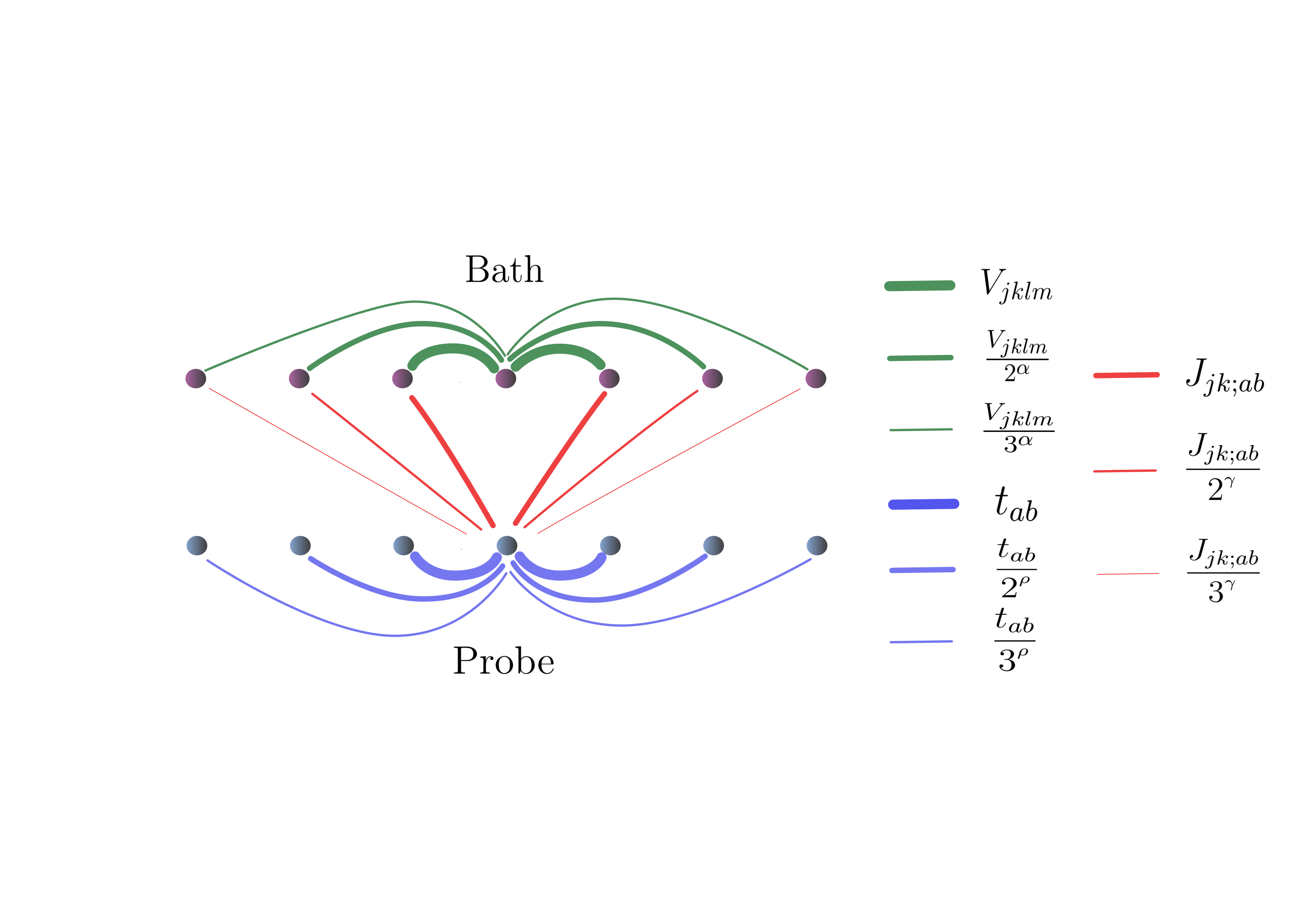}
	\caption{\textit{A pictorial representation of the bath-probe system. The upper and lower arrays of dots describe the bath and the probe respectively. Various probe sites interact among themselves through all to all random hopping \refb{hprobe1d}, depicted by the blue lines. The hopping strengths decay with distance, as represented by thinning of the blue lines. Bath probe interaction is represented by red lines. Each probe site interacts with every other bath site through a four-Fermi interaction, which is quadratic both in bath and probe fermions \refb{Hint1d}. Bath-probe interaction too decays with distance, as represented by gradual thinning of the red lines. The strength of the interactions also depend on the values of the parameter $\alpha, \rho$ and $\gamma$. The interaction becomes more and more short range for larger values of them.}}
	\label{fig:bath-probe} 
\end{figure}

We model the probe by another disordered quantum wire. The probe is smaller in the sense that each site hosts $n<N$ number of Majorana fermions of a different species $\kappa$. 
We look for a microscopic Hamiltonian that does not admit an IM phase. A free theory does the job. We allow for all to all random hopping, but the hopping strength decaying with distance. Such a system describes a Fermi liquid and can be described by the following Hamiltonian (see Fig. \ref{fig:bath-probe})
\begin{align}
	H_{probe} &= \sum_{x, x' = - \infty , \atop{ x' >x}}^\infty \sum_{1 \leq a, b \leq n} \, \frac{t^{x x'}_{ab}}{(x'-x)^\rho}  \kappa^x_a \kappa^{x'}_b  \, ,
	\label{hprobe1d}
\end{align}
where $t^{x x'}_{ab}$-s are random hoppings, satisfying  $ \overline{ t^{xx'}_{ab} } = 0 \, , \, \overline{ t^{x_1 x'_1}_{ab} t^{x_2 x'_2}_{a'b'} } = \frac{t_0^2}{n} \delta_{x_1 x_2} \delta_{x'_1 x'_2} \delta_{a a'} \delta_{b b'}$.
The over-line represents disorder average as usual.
In the limit $\rho \rightarrow \infty$ we only have nearest neighbour hopping.

Since we are talking about thermalization, it is natural to consider the bath and the probe at different temperatures. However a microscopic perspective reveals difference in temperature can be traded for a difference in coupling. Phases of matter are understood in terms of renormalization group flows of the microscopic Hamiltonian. Since couplings feature in the partition function only through the combination ``couplings times inverse temperature'', what flows is this combination. This can either be thought of as flow of coupling at fixed temperature or flow of temperature at fixed coupling. E.g. consider Ising model in triangular lattice. Its ferromagnetic (paramagnetic) phase can be realized either as low (high) temperature phase or as strong (weak) coupling phase. In the following we shall be considering the bath probe system at same temperature.

%%%%%%%%%%%%%%%%%%%%%%%%%%%%%%%%%%%% 
\subsection{Bath-probe interaction}\label{ssint}
%%%%%%%%%%%%%%%% 
The dynamics of the bath is heavily dependent on its symmetries. If the interaction with the probe, no matter how weak, upsets the symmetries the mean field solution describing the bath will be modified. Thus to avoid this situation, we make sure that the interaction does not upset the symmetries. However note that this choice of interaction doesn't make the system obviously solvable. Whereas somewhat constraining, this still leaves some room for maneuver. In particular we are able allow for all to all interaction with varying strength. Small deviations still uphold the broad results, over a sufficiently large temperature window, as we will argue later. 

Specifically, we connect this probe with the bath via a random four fermion interaction (see Fig. \ref{fig:bath-probe})
\begin{align}
	H_{int} &=  \sum_{x,x' = - \infty, \atop{x'>x}}^\infty 
	\sum_{1\leq j <k \leq N; \atop 1 \leq a< b \leq n} J^{(xx')}_{jk;ab} |x-x'|^{-\gamma} \chi^x_j \chi^x_k \kappa^{x'}_a \kappa^{x'}_b \, , 
	\label{Hint1d}
\end{align}
with $\overline{J^{(xx')}_{jk;ab} } = 0$, $\overline{J^{(x_1 x'_1)}_{jk;ab} J^{(x_2 x'_2)}_{j'k';a'b'} } = \frac{2 J_0^2}{n N^2} \delta_{x_1 x_2} \delta_{x'_1 x'_2} \delta_{a a'} \delta_{b b'}\delta_{j j'} \delta_{k k'}$.
One can add an onsite interaction term as well, without much effect.

The Hamiltonian of the combined bath-probe system is 
\begin{equation}
	H_{total}=H_{bath}+ H_{probe}+H_{int} \, ,
\end{equation}
with $H_{bath}, H_{probe}, H_{int}$ respectively given in \refb{hbath1d}, \refb{hprobe1d} and \refb{Hint1d}.

Upon disorder averaging, one finds the effective fields of the theory to be the bilocal fields $\widetilde{G}^x_\chi (\tau_1, \tau_2)$,
$\widetilde{G}^x_\kappa (\tau_1, \tau_2) := \frac{1}{n} \sum_{a=1}^n \kappa^x_a (\tau_1) \kappa^x_a(\tau_2)$ and corresponding Langrange multiplier fields $\widetilde{\Sigma}_\chi$ and $\widetilde{\Sigma}_\kappa$, whose dynamics is captured by the following action

\begin{align}
	\nonumber
	S_{eff}[\widetilde{G}_\chi, \widetilde{G}_\kappa, \widetilde{\Sigma}_\chi,\widetilde{\Sigma}_\kappa] &= N \sum_{x=-\infty}^\infty \Bigg[ -  \log \Pf (\partial_\tau - \widetilde{\Sigma}_\chi^x) - \epsilon \log \Pf (\partial_\tau - \widetilde{\Sigma}_\kappa^x) \\ \nn
	&\qquad  \qquad \quad+ \frac{1}{2} \int_0^\beta \int_0^\beta d\tau d\tau' \Bigg\{ \widetilde{\Sigma}_\chi^x(\tau, \tau') \widetilde{G}_\chi^x(\tau,\tau') \\
	\nn
	&\qquad  \qquad \quad -  \frac{V_0^2}{4} \sum_{w =1 }^\infty w^{-2\alpha} \left( \widetilde{G}_\chi^x(\tau, \tau')\right)^2 \left( \widetilde{G}_\chi^{x+w}(\tau, \tau')\right)^2 \\
	&\qquad  \qquad \quad - \frac{\epsilon t_0^2}{2} \sum_{w=1}^\infty \, w^{-2\rho} \widetilde{G}^x_\kappa(\tau_1, \tau_2) \widetilde{G}^{x+w}_\kappa(\tau_1, \tau_2) \nn \\
	&\qquad  \qquad \quad - \frac{\epsilon J_0^2}{2} \sum_{w= 1}^\infty w^{-2\gamma} \left( \widetilde{G}_\chi^x(\tau, \tau')\right)^2 \left( \widetilde{G}_\kappa^{x+w}(\tau, \tau')\right)^2  \Bigg\}  \Bigg] \, .
	\label{seff}
\end{align}

It is instructive to first study the somewhat trivial case of $t_0=0$. Assuming translational invariance, in this case we have the saddle point equations
\begin{align}
	\left( V^2 \widetilde{G}_\chi^3 + \epsilon J^2 \widetilde{G}_\chi \widetilde{G}_\kappa^2 \right) \circ \widetilde{G}_\chi   &= - 1 \, , \,
	J^2 \widetilde{G}_\chi^2 \widetilde{G}_\kappa \circ \widetilde{G}_\kappa = - 1 \, , \label{SDcombo}
\end{align}
in deep infrared, with $V^2 =  V_0^2 \zeta(2\alpha), J^2 =  J_0^2 \zeta(2\gamma)$ and $\epsilon=n/N$. 
Their claim to fame is that they continue to enjoy the time reparameterization symmetry, with both $\widetilde{G}_\chi$ and $\widetilde{G}_\kappa$ transforming as in \refb{rep}. 
The homogeneous appearance of $\widetilde{G}_\chi$ and $\widetilde{G}_\kappa$, suggests the following ansatz 
\begin{align}\label{eq:ansatz-probe+bath}
	G_{s,\kappa} &= A G_{s,\chi} \, ,
\end{align}
Consistency of the two saddle point equations fixes
\begin{align}
	A &=  \frac{V}{J (1-\epsilon)^{1/2}} \, .
	\label{Aval}
\end{align}
Thus one has the saddle point configurations
\begin{align}
	\nn
	G_{s, \chi} (\tau_1, \tau_2) &= \left( \frac{1-\epsilon}{4\pi V^2} \right)^{1/4} \left[ \frac{\pi}{\beta \sin \frac{\pi \tau_{12}}{\beta}}\right]^{1/2} \sgn{} \tau_{12} \, , \,\\
	G_{s, \kappa} (\tau_1, \tau_2) &= \frac{V^{1/2}}{J (1-\epsilon)^{1/4} (4\pi)^{1/4}} \left[ \frac{\pi}{\beta \sin \frac{\pi \tau_{12}}{\beta}}\right]^{1/2} \sgn{} \tau_{12} \, ,
	\label{Gsaddlecombo}
\end{align}
which spontaneously break the reparameterization symmetry down to $SL(2, \mathbb{R})$. 
The upshot of the above analysis is that connecting the probe with the bath does not affect the symmetry of the bath and its spontaneous breaking.
%%%

To see this more clearly we expand the action \refb{seff} around the saddle \refb{Gsaddlecombo} in terms of fluctuation fields $g_\chi, g_\kappa, \sigma_\chi, \sigma_\kappa$ \refb{fluc}, defined through
\begin{align}
	\nn
	\widetilde{G}^x_{\chi} &= G_{s,\chi} + |G_{s,\chi}|^{-1} g^x_\chi \, ,\\
	\nn
	\widetilde{G}^x_{\kappa} &= G_{s,\kappa} + |G_{s,\chi}|^{-1} g^x_\kappa \, ,\\
	\nn
	\widetilde{\Sigma}^x_{\chi} &= \Sigma_{s,\chi} + |G_{s,\chi} |\  \sigma^x_\chi \, ,\\
	\widetilde{\Sigma}^x_{\kappa} &= \Sigma_{s,\kappa} + |G_{s,\chi} |\  \sigma^x_\kappa \, .
	\label{fluc}
\end{align}

Further integrating $\sigma_\chi, \sigma_\kappa$ out, we obtain the following action

\begin{align}
	S^{total}_{eff} &= \frac{N}{8\pi} \int dp~\Bigg[ g_\chi(p) \circ R_\chi(p) \circ g_\chi(-p) 
	+ g_\kappa(p) \circ R_\kappa(p) \circ g_\kappa(-p) \\ 
	&\qquad  \qquad \qquad + g_\kappa(p) \circ R_{\kappa \chi}(p) \circ g_\chi(-p) \Bigg] \, , \nn\\~\nn \\
	\nn
	\text{with,} \, \, R_\chi(p) 
	&= 3V^2 \Big \{ \frac{1}{1-\epsilon} \left( K_c^{-1} - 1 + \frac{2\epsilon}{3}  \right) 
	+ \frac{V_0^2}{3V^2} \left[ 2\zeta(2\alpha) - Li_{2\alpha}(e^{ip}) - Li_{2\alpha}(e^{-ip})\right] \Big\} \, , \\
	\nn
	R_\kappa(p) 
	&=  3 \epsilon J^2 \left[ K_c^{-1} - 1/3 \right] \, , \\
	R_{\kappa \chi}(p) 
	&= - \frac{2 \epsilon VJ}{(1-\epsilon)^{1/2}}   \frac{ Li_{2\gamma}(e^{ip}) + Li_{2\gamma}(e^{-ip} )  }{ \zeta(2\gamma)}  \, . \label{seffgcomb}
\end{align}

Since we are primarily interested in the fate of the probe, we integrate $g_\chi$ out to obtain an effective action for the probe exclusively: 
%\begin{widetext}
\begin{align}
	&S_{eff}^{probe} = \frac{N}{8 \pi} \int dp\, g_\kappa(p) \circ R'_\kappa \circ g_\kappa(-p) \, , 
	\label{Rpk}
\end{align}
where $R'_\kappa = - \frac{R_{\kappa \chi}^2(p)}{4 R_\chi(p)} \, + \,  R_{\kappa} $.
The rather clumsy expression of $R'_\kappa$ has the nice limit
\begin{align}
	\lim_{p \rightarrow 0} \lim_{K_c \rightarrow 1} R'_\kappa &= - \frac{16 \epsilon^2 V^2 J^2}{1-\epsilon} \times \frac{1-\epsilon}{8 \epsilon V^2} + 2 \epsilon J^2 = 0 \, ,
\end{align}
which entails that the effective description of the probe continues to have the same (pseudo)Goldstones or soft modes as the bath. This implies that the effective dynamics of the probe too is dominated by the soft modes, which calls for restricting $S_{eff}^{probe}$ \eqref{Rpk} to these modes.
To this end, we expand $R'_\kappa$ in $p^2$ and $(K_c^{-1}-1)$, and get
\begin{align}
	\nn
	R'_\kappa &= 
	\frac{J^2}{4} \big[ 3(1+\epsilon) (K_c^{-1}-1) \\
	&+
	\left( (1 - \epsilon) \frac{\zeta(2\alpha-1) }{\zeta(2\alpha)} + 2 \epsilon \frac{\zeta(2\gamma-1) }{\zeta(2\gamma)} \right) p^2
	\big] + \dots
	\label{B9}
	\, , 
\end{align}
to leading order.

Up to a multiplicative factor of $A^{-2}$, this expression agrees with $R_\chi$ in \eqref{seffgcomb}, to leading order in $\epsilon$ and quadratic order in $p$. Further noting that for soft modes, $g_\chi, \, g_\kappa$ are  proportional to reparameterizations of $G_{s,\chi}$ and $G_{s,\kappa}$ respectively \eqref{fluc} and then using \eqref{eq:ansatz-probe+bath}, we find $g_\kappa = A g_\chi$ for soft modes. Putting this in the probe effective action \eqref{Rpk}, we see the soft mode sector of it coincides with that of the bath to leading order in $\epsilon$. Thus after taking $\mathcal{O} (1/\beta \mathcal{V})$ corrections into account, we find the soft mode action for the probe to be same as that of the bath \refb{ssoftbpsmall}. 
This immediately implies that the dominant piece in probe four point function has the same form as \refb{Fbigbath}. Hence the probe has the identical diffusion coefficient and Lyapunov spectrum as the bath. In particular, the probe has turned into an IM.  For finite $\epsilon$, the probe back-reaction renormalizes the bath, leading to only quantitative changes. The bath still continues to induce diffusion coefficient and Lyapunov spectrum, and consequently its phase, onto the probe. 
Thus this bath-probe system constitutes an example of {\it phase thermalization}. It is worthwhile to note that we loosely refer to this phenomenon as phase transition but further exploration is needed to understand the nature of this `change of phase'. In fact since no obvious symmetry is broken it is likely to be a crossover %\footnote{We thank the anonymous referee for pointing out this possibility and drawing our attention to \cite{Altland_2019}.}
from Fermi liquid to IM phase, as advocated in \cite{Altland_2019}.
%%%

Now we turn on finite but small $t_0$. Ignoring the back reactions of the probe, we get an action for $\widetilde{G}_\kappa$. Upon assuming translational invariance, this action  coincides (upon appropriate identification of parameters) with that of a system with only one species of fermions, interacting through random hopping of strength $t_0$ and random quartic interactions of strength $U_0$, when expanded around the saddle for vanishing hopping. For the latter system it is known that above a transition/crossover temperature $T_c \sim t_0^2/U_0$, IM phase persists even after turning on hopping \cite{Song:2017pfw}. We choose to work with large enough quartic couplings compared to the hopping, so that this temperature is fairly small. Back reaction of the probe does not alter the above conclusions, since it modifies the effective action only quantitatively. 

Turning on other random interactions also keeps the conclusions unaltered. The only relevant perturbation would be a bilinear term, which is prohibited by $\kappa \rightarrow - \kappa$ and $\chi \rightarrow -\chi$ symmetry. Other random q-Fermi interactions are irrelevant and hence would matter only above some high transition/crossover temperature. Dimensional analysis suggests this transition/crossover temperature should scale as $(U_0^q/c_q^4)^{1/(4-q)}$, where $c_q$ is the coupling strength of q-Fermi  
interaction. More realistic analysis must include interaction with phonons, although we shall not attempt it here.

The similarity between $H_{int}$ and $H_{bath}$ may raise doubt whether it is primarily $H_{int}$ which is turning the probe into an IM, rather than the bath itself. The diffusion coefficient of the probe being independent of the bath-probe interaction strength $J_0$ to leading order in $\epsilon$, confirms that the specific form of $H_{int}$ is not of key importance.

%%%%%%%%%%%%%%%%%%%%%%%%%%%%%%%%%%%%%%%%%%%%%%%%%%%%%%%%%%%%%%%%%
\section{Discussions}  \label{sdisc}
%%%%%%%%%%%%%%%%%
Remarkably, the bath-probe system is solvable for any $\epsilon < 1$, \emph{i.e.} even when the probe is comparable to the bath in size. 
Since the bath-probe system can not be thought of as a larger SYK chain (due to absence of a quartic term in $\kappa$ fermions as well as $\alpha$ and $\gamma$ being independent) this is rather non-trivial. The curious case of $\epsilon=1$, i.e. bath and probe having same size, seems to be singular and deserves further explorations. The phase structure in $\alpha, \gamma$ plane also calls for more scrutiny. 

Strictly speaking our analysis is valid for $\alpha >1$. However using the analytic continuation of $\text{Li}_s(z)$ function we can extend its range of validity even beyond $\alpha = 1$. As we speculated above  (see Figure. \ref{fig:lambda-for-alpha-less-than-one}), the $0 < \alpha < 1$ regime can potentially encapsulate intriguing physics.  The Lyapunov exponent in that regime takes qualitatively different values namely, positive, zero and negative values which are usually associated to systems with chaotic, conservative and dissipative dynamics respectively. We stress again this observation has to be taken with a grain of salt but definitely asks for further future investigation. 

Experimental verification of phase thermalisation in the context of this particular model might be possible along the lines of \cite{chew2017approximating, Pikulin:2017mhj, Danshita:2016xbo}. Especially promising is the proposal of \cite{chew2017approximating} realizing SYK model as a quantum dot with Majorana wires plugged in. An SYK lattice can be realized as a lattice of such quantum dots. Our model for the bath would further require some interaction between these quantum dots.

The most alluring problem however is to develop a general framework for the notion of phase thermalization. Finding more examples of this phenomenon would make some headway.

%%%%%%%%%%%%%
\acknowledgments{We would like to thank Debmalya Chakaraborty and Saheli Sarkar for useful comments. We are also grateful to the referee whose valuable suggestions significantly improved the content of the paper. PB would like to acknowledge the support provided by the Max Planck Partner Group grant MAXPLA/PHY/2018577, Fulbright Foundation (Award No. 2564/FNPDR/2020) and Ramanujan fellowship grant RJF/2023/000007 from the Anusandhan National Research Foundation (ANRF), India. Research of BC is supported by ST/T000775/1 grant. Research of SM is supported by Laureate Award 15175 ``Modularity in Quantum Field Theory and Gravity'' of the Irish Research Council. This work was initiated during our stay at ICTS-TIFR, Bangalore. }
%%%%%%%%%%%%%%
%\pagebreak
\appendix

%%%%%%%%%%%%%%%%%%%% %%%%%%%%%%%%%%%%%%%% %%%%%%%%%%%%%%%%%%%% 
\section{Effective dynamics of soft modes of the Incoherent Metallic bath} \label{appA}
For the incoherent metallic (IM) bath, described by the Hamiltonian in \refb{hbath1d}, the 
disorder averaged theory can be expressed as a path integral over the bilocal fields $\widetilde{G}, \widetilde{\Sigma}$ with the action

 \begin{align}
	& S_{eff}[\widetilde{G}, \widetilde{\Sigma} ] = N \sum_{x = -\infty }^\infty \bigg[ 
	-\log \Pf{} (\partial_\tau - \widetilde{\Sigma}^x) + \frac{1}{2} \int_0^\beta d\tau_1 d\tau_2
	\bigg\{ 
	\widetilde{\Sigma}^x(\tau_1, \tau_2) \widetilde{G}^x(\tau_1, \tau_2)\nn \\
	& \qquad  \qquad \qquad \qquad \qquad- \frac{V_0^2}{4} \sum_{w=1 }^\infty w^{-2\alpha} \widetilde{G}^x(\tau_1,\tau_2)^2 \widetilde{G}^{x+w}(\tau_1, \tau_2)^2  \bigg\} 
	\bigg] 
	\, . \label{Sbath}
\end{align}

We expand $\widetilde{G}^x, \widetilde{\Sigma}^x$ around their saddle point values $G_s, \Sigma_s$ as follows
\begin{align}
	\nn
	\widetilde{G}^x &= G_s + |G_s|^{-1} g^x \, , \\
	\widetilde{\Sigma}^x &= \Sigma_s + |G_s| \, \sigma^x .
	\label{sadval}
\end{align}
Further the effective action in \refb{Sbath} is expanded around the saddle configuration \refb{Gsaddle1}, to quadratic order in $g^x, \sigma^x$, and then we integrate $\sigma^x$ out. Going to the momentum space with the following convention 
\begin{align}
	\phi(x) &= \frac{1}{2\pi} \int_0^{2\pi} dp \, e^{-ipx} \phi(p) \, , 
\end{align} we obtain the following effective action governing the fluctuations $g(p)$
	\begin{align}
		S_{eff}[\, g \, ] &= \frac{N}{8\pi} \int_0^{2\pi} dp \, g (p) \circ \left[ 3 V_0^2 \zeta(2\alpha) \left( K_c^{-1} -1 \right) 
		+ V_0^2 \left[ 2 \zeta (2\alpha) - Li_{2\alpha} (e^{ip})  - Li_{2\alpha} (e^{-ip}) \right] \right]  \circ g(-p) \, , \label{eqnA4}
	\end{align} 
with the kernel $K_c$ defined in \refb{kcdefn}. $\zeta(2\alpha)$ and $Li_{2\alpha}$ are respectively Riemann zeta function and polylogarithm function.

Soft modes are the changes in $G_s$ under reparameterizations. Any function $\varepsilon(p)$ on the Euclidean time circle defines a reparameterization
\begin{align}
	g(p) &= 
	|G_{s}| \delta_{\varepsilon(p)} \widetilde{G}(p).
\end{align} 
They can conveniently be expanded in Fourier modes $g_m$ as defined in \refb{eqn8} of the main text. 
These modes satisfy normalization relations
\begin{align}
	g_m \circ g_{m'} &= \delta_{m+m'} \frac{b^4}{16} \left( \frac{2\pi}{\beta} \right)^4 \beta^2 \frac{|m| (m^2 -1)}{3}\ ,
\end{align} where $b= \left( 4 \pi V_0^2 \zeta(2\alpha) \right)^{1/4}$.
Note that $g_0, \, g_{\pm 1}$ have vanishing norm. These modes correspond to global reparameterizations and are not soft modes. This is why they have been excluded in the Fourier expansion \refb{gpfourier} of $g(p)$.

Strictly speaking, only $p=0$ modes in $K_c=1$ sector are soft modes. However they are continuously connected to modes with non-zero momenta in the same sector. So it makes sense to restrict to $K_c=1$ and allow all momenta. By abuse of notation, we shall refer to all such modes as soft modes. 
The effective action for $K_c=1$ sector is 

\begin{align}
	S_{eff}^{soft} 
	&=
	\frac{N \alpha_K}{ 256 \pi \mathcal{V}} \left( \frac{2\pi}{\beta}\right)^2 \int_0^{2\pi} dp \, \sum_{|m| \geq 2} \varepsilon_m(p) \varepsilon_{-m}(-p) |m| (m^2-1)
	\bigg[ 
	\frac{ 2\pi |m|}{\beta} \nn \\
	&\qquad \qquad \qquad \qquad \qquad \qquad + \frac{2\pi \mathcal{V}}{\alpha_K} \times \frac{2 \zeta(2\alpha) - Li_{2\alpha}(e^{ip}) - Li_{2\alpha}(e^{-ip}) }{3 \zeta(2\alpha)}
	\bigg] \, .
	\label{ssofteff}
\end{align}

For small momenta or equivalently long wavelengths one can approximate
\begin{align}
	\zeta(2\alpha) - \frac{Li_{2\alpha}(e^{ip}) + Li_{2\alpha}(e^{-ip})}{2} \sim \frac{1}{2} \zeta(2\alpha -1) p^2  \,.
\end{align}
In this regime, \eqref{ssofteff} reduces to \refb{ssoftbpsmall}. 

Contribution of these modes to the four point function $\mathcal{F}$ in \refb{Fcon} is 
\begin{align}
	\nn
	\mathcal{F}^{big}(p;1,2,3,4) 
	=\, &N  \sum_{m} \langle \varepsilon_m(p) \varepsilon_{-m}(-p) \rangle \\
	& \delta_m G_s^\chi(p;1,2) \delta_{-m} G_s^\chi (-p;3,4)\ .
	\label{4pt}
\end{align} 
Upon using 
\begin{align}
	\nn
	\langle \varepsilon_m(p) \varepsilon_{-m}(-p) \rangle =\,  &\frac{32 \mathcal{V}}{N V^2 b^4 \alpha_K |m| (m^2-1)} \\
	\nn
	&\left( \frac{\beta}{2\pi}\right)^2 \frac{1}{\omega_m + D_{\alpha} p^2} ,
	\label{fluc2ptcorrel} \\
	\delta_n G_s (\tau_{12}) 
	=\, \frac{i b^{2}}{4 } \left( \frac{2\pi}{\beta}\right)^2 &\frac{e^{-2i\pi ny_{12}/\beta}}{\sin\frac{\pi \tau_{12}}{\beta}} G_s(\tau_{12})^{-1} f_n(\tau_{12}) \, ,  
\end{align}
with $D_{\alpha}$ given in \refb{diffbath}, we get
	\begin{align}
		\frac{\mathcal{F}^{big}(p;\tau_1, \tau_2;\tau_3,\tau_4)}{G_s(\tau_{12}) G_s(\tau_{34})} &=\frac{32 \pi \mathcal{V} }{\alpha_K} 
		\sum_{|m| \geq 2} \frac{1}{|m| (m^2-1)} \frac{ e^{-2\pi i (y_{12} - y_{34})/\beta}}{|\omega_m| + D_{\alpha} p^2} f_m(\tau_{12})  f_m(\tau_{34}) \, . 
		\label{eqnA11}
	\end{align}
Since $|\omega_m| \sim 1/\beta$ and $p^2$ is small, this is an $\mathcal{O} (\beta \mathcal{V})$ contribution and towers over $\mathcal{O}(1)$ contributions coming from other modes. 

%%%%%%%%%%%%%%%%%%%

\bibliographystyle{bibstyle}
\bibliography{APS2}

\end{document}